\documentclass[conference]{IEEEtran}

\usepackage{hhline}
\usepackage{bm}
\usepackage{color}
\usepackage{longtable}	
\usepackage{multicol}
\usepackage{float}
\usepackage{multirow}
\usepackage{subfigure}
\usepackage{cite}
\usepackage{amsmath,amssymb,amsfonts}
\usepackage{algorithmic}
\usepackage{graphicx}
\usepackage{textcomp}
\usepackage{xcolor}
\def\BibTeX{{\rm B\kern-.05em{\sc i\kern-.025em b}\kern-.08em
    T\kern-.1667em\lower.7ex\hbox{E}\kern-.125emX}}
\IEEEoverridecommandlockouts

 \pdfoutput=1

\usepackage{atbegshi}
\AtBeginDocument{\AtBeginShipoutNext{\AtBeginShipoutDiscard}}

\newcommand{\E}{\mathbb{E}}

\newcommand{\Var}{\mathrm{Var}}

\begin{document}

\title{Investigating the Impacts of Stochastic Load Fluctuation on Dynamic Voltage Stability Margin Using Bifurcation Theory}

\author{\IEEEauthorblockN{Georgia Pierrou}
\IEEEauthorblockA{\textit{Department of Electrical and Computer Engineering} \\
\textit{McGill University}\\
Montreal, QC H3A 0G4, Canada \\
georgia.pierrou@mail.mcgill.ca}
\and
\IEEEauthorblockN{Xiaozhe Wang}
\IEEEauthorblockA{\textit{Department of Electrical and Computer Engineering} \\
\textit{McGill University}\\
Montreal, QC H3A 0G4, Canada\\
xiaozhe.wang2@mcgill.ca}
}

\thanks{This work is supported by Natural Sciences and Engineering Research Council (NSERC) under Discovery Grant NSERC RGPIN-2016-04570 and by the Stavros S. Niarchos Foundation McGill Fellowship.
}


\maketitle

\begin{abstract}
This paper studies the impacts of stochastic load fluctuations, namely the fluctuation intensity and the changing speed of load power, on the size of the voltage stability margin. To this end, Stochastic Differential-Algebraic Equations (SDAEs) are used to model the stochastic load variation; bifurcation analysis is carried out to explain the influence of stochasticity. Numerical study and Monte Carlo simulations on the IEEE 14-bus system demonstrate that a larger fluctuation intensity or a slower load power changing speed may lead to a smaller voltage stability margin. Particularly, this work may represent the first attempt to reveal the influence of the time evolution property of the driving parameters on the voltage stability margin in power systems.
\end{abstract}



\section{Introduction}

The growing load demand along with the integration of renewable energy sources (RES) have posed great challenges to 
the stability assessment of power systems \cite{Haesen09} -\cite{Qiu08}.
To incorporate the randomness, stochastic power system dynamic models have been proposed in \cite{Milano13} -\cite{Wangxz15} for power system voltage stability study. It has been shown in \cite{Wang17} that although the conventional deterministic model can well approximate the stochastic model under some mild conditions, there are critical cases in which it may fail to provide accurate results in the voltage stability assessment. To deal with these challenges, it is important to conduct systematic study to assess the impacts of the randomness on power system voltage stability.

Bifurcation theory has been widely used to explain the dynamic mechanisms of voltage collapse \cite{Chiang90} -\cite{Canizares95}. However, the randomness brought about by the loads and RES was not considered in these studies.
To incorporate the randomness into dynamic voltage stability, the authors of \cite{DeMarco87} considered load fluctuations in voltage stability study and derived the first exit time as a measure for voltage stability assessment. Similar approaches were adopted in \cite{Nwankpa93}, \cite{Nwankpa2000} to study the intensities of load fluctuations and the uncertainty of wind power on system voltage stability.
The authors of \cite{Qiu08} used the bifurcation theory to analyze how a stochastic load model 
may affect the voltage profile. However, the conclusions are based on a single realization of the stochastic dynamic system.

In this paper, we strive to systematically investigate the impacts of stochastic load fluctuations on the dynamic voltage stability margin. Inspired by the bifurcation theory and the analytical results in \cite{Wangxz15}, we focus on the influence of two sets of parameters, namely, the intensity of load fluctuation and the changing speed of load power. Detailed numerical study and Monte Carlo simulations are conducted on the IEEE 14-bus system. 
Interestingly, we find that the changing speed of load power (i.e., the time evolution property of the driving parameters) may affect the size of the voltage stability margin. 
Such interesting results have not been reported in the previous literature. From the extensive simulation study it can be concluded that the randomness of load fluctuations should be incorporated into the dynamic voltage stability study and the time evolution property needs to be carefully considered in order to achieve an accurate stability assessment.

\section{The Stochastic Power System Model}\label{1}
\vspace{-5pt}
 To study the influence of the randomness on the dynamic behavior of power systems, 
the stochasticity can be incorporated in the classic power system model represented by a set of Differential-Algebraic Equations (DAEs) as follows:
\begin{equation}
\label{eq:sdae_model}
\begin{gathered}
\dot{\bm{x}} = \bm{f}(\bm{x},\bm{y},\bm{p}, \bm{\eta}) \\
\bm{0}=\bm{g}(\bm{x},\bm{y}, \bm{p})
\end{gathered}
\end{equation}
where $\bm{x}$ is the vector of state variables, e.g., generator angles and speeds, states of dynamic loads; $\bm{y}$ is the vector of algebraic variables, e.g., bus voltage magnitudes and phases;
$\bm{p}$ is the vector of parameters, e.g., load powers;
$\bm{\eta}$ is the vector of stochastic perturbations describing, for instance, load fluctuations and renewable energy variations. $\bm{f}$ 
describe the behavior of the dynamic components, including generators and their control, dynamic loads etc.; $\bm{g}$ 
describe the static behaviors and network power flow constraints.


In this paper, we focus on the randomness brought about by load fluctuations, which can be described 
by a vector Ornstein-Uhlenbeck  process as in previous work \cite{Milano13}, \cite{Ghanavati16}:
%
\begin{equation}
\label{eq:ouprocess}
\dot{\bm{\eta}} = -A_{\bm{\eta}}{\bm{\eta}} + \sigma B_{\bm{\eta}}{\bm{\xi}}, \quad t \in [0,T]
\end{equation}
where $A_{\bm{\eta}}=\mbox{diag}([\alpha_{1},...,\alpha_{k}])$ and is positive definite; $\sigma$ describes the intensity of stochastic perturbations;  $B_{\bm{\eta}}=\mbox{diag}([\beta_{1},...,\beta_{k}])$ denotes the relative strength between perturbations. $\bm{\xi}\in \mathbb{R}^k$ is a vector of independent Gaussian random variables described by $\xi_{i}=\frac{d{W}_{ti}}{dt}$, where $W_{ti}$ is a Wiener process, $i \in \{1,...,k\}$.

For an initial condition $\eta_i(0) \sim \mathcal{N}(0,(\sigma\beta_i)^2/2\alpha_i)$, $i \in \{1,...,k\}$, the stochastic process $\eta_i$ has the following statistical characteristics \cite{Gardiner}:
\begin{itemize}

\item $\E[\eta_i(t)]=0, \quad  \forall t \in [0,T],$
\item $\Var[\eta_i(t)]=(\sigma\beta_i)^2/2\alpha_i, \quad \forall t \in [0,T],$

\item $\mbox{Aut}[\eta_i(t_{p}),\eta_{i}(t_{q})] = e^{-\alpha |t_{q}-t_{p}|}, \quad \forall t_{p},t_{q} \in [0,T]$.

\end{itemize}
Therefore, ${\bm{\eta}}$ is a vector of stationary autocorrelated Gaussian processes describing the load fluctuations.

If the algebraic Jacobian matrix $g_{\bm{y}}$ is non-singular, which is typically satisfied in normal operating state, $\bm{y}$ can be expressed by $\bm{x}$ and $\bm{p}$. Therefore, the stochastic power system model (\ref{eq:sdae_model})-(\ref{eq:ouprocess}) can be represented by:
\begin{eqnarray}
\label{eq:sde_model_withouty_1}
\dot{\bm{x}} &=& \bm{H}(\bm{x},\bm{p}, \bm{\eta}) \\
\dot{\bm{\eta}} &=& -A_{\bm{\eta}}{\bm{\eta}} +\sigma B_{\bm{\eta}}{\bm{\xi}}
\label{eq:sde_model_withouty_2}
\end{eqnarray}
Let $\bm{u}=\begin{bmatrix} {\bm{x}}, {\bm{\eta}}
\end{bmatrix}^T$ and $B=\begin{bmatrix} \bm{0}, B_{\bm{\eta}} \end{bmatrix}^T$,  (\ref{eq:sde_model_withouty_1})-(\ref{eq:sde_model_withouty_2}) is equivalent to:
\begin{equation}
\label{eq:sde_model_u}
\dot{\bm{u}} = \bm{G}(\bm{u},\bm{p})+\sigma B \bm{\xi} \\
\end{equation}

\subsection{The Stochastic Load Model}
In this paper, we consider the dynamics of the exponential recovery load (ERL) model that play a significant role in voltage stability analysis. As a result, the detailed expression of (\ref{eq:sdae_model}) takes the following form (\ref{eq:erload})-(\ref{eq:erl_power}): 
\begin{equation}
\label{eq:erload}
\begin{gathered}
\dot{x}_p=-x_p/T_p+p_s-p_t \\
\dot{x}_q=-x_q/T_q+q_s-q_t
\end{gathered}
\end{equation}
where ${x}_{p}$ and ${x}_{q}$ are related to the active and reactive power dynamics; ${T_{p}}$ and ${T_{q}}$ are the corresponding power time constants; ${{p}_{s}}$ and ${{p}_{t}}$ are the static and transient real power absorption; ${{q}_{s}}$ and ${{q}_{t}}$ are similarly defined as for the reactive power absorption. 
If the randomness of load fluctuation is incorporated into the dynamic load model, similar to the approach in \cite{Milano13}, \cite{Perninge11}, \cite{Wangxz:2017}, we have the following relationships:  
\begin{equation}
\label{eq:exp_rec_load_sde}
\begin{gathered}
{{p}_{s}}={({p}_{0}+\eta_{p}(t))}(\frac{V}{{V}_{0}})^{{\alpha}_{s}} \quad {{p}_{t}}={({p}_{0}+\eta_{p}(t))}(\frac{V}{{V}_{0}})^{{\alpha}_{t}} \\
{{q}_{s}}={({q}_{0}+\eta_{q}(t))}(\frac{V}{{V}_{0}})^{{\beta}_{s}} \quad
{{q}_{t}}={({q}_{0}+\eta_{q}(t))}(\frac{V}{{V}_{0}})^{{\beta}_{t}} \\
\end{gathered}
\end{equation}
where ${{p}_{0}}$ and ${{q}_{0}}$ are the nominal active and reactive power of the ERL; $\eta_p$ and $\eta_q$ are Ornstein-Uhlenbeck processes (see (\ref{eq:ouprocess})) describing the stochastic perturbations around the nominal power values; ${{\alpha}_{s}}$, ${{\beta}_{s}}$, ${{\alpha}_{t}}$ and ${{\beta}_{t}}$ are exponents related to the steady state load response 
and the transient load response, respectively; ${{V}_{0}}$ is the nominal bus voltage. 
The resulting active and reactive power of the ERL are given by:
\begin{equation}
\label{eq:erl_power}
\begin{gathered}
p=x_p/T_p+p_t \\
q=x_q/T_q+q_t
\end{gathered}
\end{equation}

It should be noted that similar procedures can be applied to other dynamic load models such as the frequency-dependent load model, the thermostatic recovery load model, etc. to include the stochastic perturbations into their nominal powers.

\section{The Impacts of Randomness on Voltage Stability Margin} \label{3}
\subsection{Bifurcation Theory}
Bifurcation theory has been widely used in the literature to explain 
voltage stability in power systems \cite{Chiang90} -\cite{Canizares95}. Generally, we consider a nonlinear dynamic system described by:
\begin{equation}
\label{eq:ode}
\bm{\dot{x}}  = \bm{F}(\bm{x},\bm{p}(\varepsilon t))
\end{equation}
where $\bm{x}\in \mathbb{R}^n$ are the state variables and $\bm{p}$ are the slowly changing parameters. In power system voltage stability study, such parameters are typically the real and/or reactive power of loads and renewable generators.
%
%
As the parameters $\bm{p}$ vary, a bifurcation may occur leading to a qualitative change in the behavior of the system, such as the change of stability of the equilibrium point, the emergence of oscillations
and even the disappearance of equilibrium points. 


The Saddle-Node Bifurcation (SNB) is typically used to explain the dynamic mechanism of voltage collapse.
As the parameters $\bm{p}$, e.g., load powers, change, the equilibrium point $\bm{x^\star}$ will vary in the state space leading to a slow decrease in voltage magnitudes. At the critical load power $\bm{p_1}$, the voltage magnitudes sharply decrease and the system loses stability by $\bm{x^\star}$ disappearing in a SNB. The difference between the power at the current operating point and the power at the SNB point is defined as the voltage stability margin of the system.
A necessary condition for the SNB is the singularity of the Jacobian matrix $F_{\bm{x}}$ \cite{Canizares92}.
If we describe (\ref{eq:ode}) in the slow timescale $s=\varepsilon t$, we have: 
\begin{equation}
\label{eq:det_ode_slow}
{\bm{x}}^{\prime}  = \frac{1}{\varepsilon} \bm{F}(\bm{x},\bm{p}(s))
\end{equation}
where $^\prime$ denotes $\frac{d}{ds}$.
Let $\cal{M}$ = $\{\bm{x}^\star(s):F(\bm{x^\star}(s))=0\}$ be the uniformly asymptotically stable slow manifold of the deterministic system (\ref{eq:det_ode_slow}). 
According to Fenichel's theorem, for sufficiently small $\varepsilon$, there exists an invariant manifold $\cal{M_{\varepsilon}}$ = $\{\bm{\bar{x}}(s,\varepsilon):\bm{\bar{x}}(s,\varepsilon)=\bm{x^\star}(s)+ \cal{O}(\varepsilon)\}$ at a distance of order $\varepsilon$ from $\cal{M}$, which attracts nearby solutions exponentially fast. In other words, the solution of (\ref{eq:det_ode_slow}) will stay in $\cal{M}_\varepsilon$  despite the initial transient. 

\vspace{-3pt}
\subsection{Concentration of Sample Path around the Stable Manifold}\label{sectionconcentrationofpaths}
\begin{figure}[!ht]
\vspace{-6pt}
\centering
\includegraphics[width=3in ,keepaspectratio=true,angle=0]{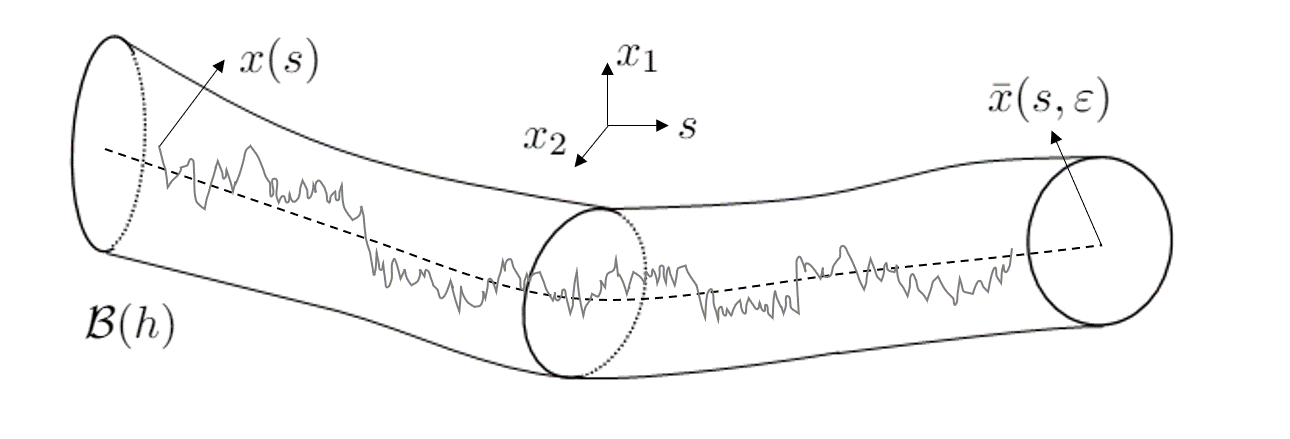}
\vspace{-5pt}
\caption{
The trajectory of the stochastic model (\ref{eq:stoch_ode_slow}) is concentrated in the neighborhood $\cal{B}$$(h)$ of the trajectory of the deterministic model (\ref{eq:det_ode_slow}) if  $h\gg\sigma$.}

\vspace{-5pt}
\label{samplepath}
\end{figure}

In the presence of white Gaussian noise, the slowly time-dependent equation (\ref{eq:ode}) becomes:
\begin{equation}
\label{eq:stoch_ode}
\dot{\bm{x}}  = \bm{F}(\bm{x},\bm{p}(\varepsilon t)) + \sigma\Sigma \bm{\xi}
\end{equation}
where 
$\sigma$
describes the intensity of the stochastic perturbations and $\Sigma$ describes the relative strength and the correlation between noises. 
If written in the slow time scale
$s=\varepsilon t$, it takes the following form: 
\begin{equation}
\label{eq:stoch_ode_slow}
{\bm{x}}^{\prime}  = \frac{1}{\varepsilon} \bm{F}(\bm{x},\bm{p}(s)) +\frac{\sigma}{\sqrt{\varepsilon}}\Sigma \bm{\xi}
\end{equation}
%
It has been shown in \cite{Wangxz15}, \cite{Gentz06} that the sample path of the stochastic dynamic model (\ref{eq:stoch_ode_slow}) is unlikely to leave the neighbourhood $\cal{B}$$(h)$ of the invariant manifold $\cal{M}_\varepsilon$ of (\ref{eq:det_ode_slow}), if $h\gg\sigma$.  
Particularly, 
$\cal{B}$$(h):=\{\bm{x}(s): \langle \bm{x}(s)-\bar{\bm{x}}(s,\varepsilon),\bar{X}(\varepsilon)^{-1}(\bm{x}(s)-\bar{\bm{x}}(s, \varepsilon))\rangle<h^2\}$,  where $\langle\rangle$ denotes the inner product and $\bar{X}(\varepsilon)$ describing the cross section of $\cal{B}$$(h)$ is well defined (See Appendix in \cite{Wangxz15}). The sample path of (\ref{eq:stoch_ode_slow}) surrounding the invariant manifold of the trajectory of the deterministic system (\ref{eq:det_ode_slow}) is illustrated in Fig. \ref{samplepath}.


The above results reveal that the trajectory of the stochastic power system model (\ref{eq:stoch_ode_slow}) will be likely concentrated in a small neighborhood of the trajectory of the deterministic power system model (\ref{eq:det_ode_slow}) when the power system is operating in normal conditions. 
 Therefore, a natural question to ask is what will happen if the system is getting close to the SNB point. What are the impacts of the randomness on SNB and thus on the voltage stability margin? 


\vspace{-5pt}
\subsection{The Impact of Noise on Saddle-Node Bifurcation}

It is revealed from the previous results that the shape of the concentration neighborhood $\cal{B}$$(h)$ is affected by $\varepsilon$ (in the detailed expression of $\bar{X}(\varepsilon)^{-1})$ while the depth of $\cal{B}$$(h)$ is determined by $\sigma$. Mathematically, $\varepsilon$ describes the decoupling between the slow and the fast dynamics,  whereas $\sigma$ describes the intensity of the stochastic perturbations.
Inspired by theses results, we intend to study the impact of the two parameters $\sigma$ and $\varepsilon$ on SNB.

In the context of power system voltage stability study, $\sigma$ describes the intensity of load fluctuations and $\varepsilon$ describes the changing speed of load power or the other parameters depending on the problem of interest. 
In this work, we assume that the real and reactive power of the constant PQ loads experience a gradual increase with respect to time:
\begin{equation}
\label{eq:PQincrease}
\begin{gathered}
P(\varepsilon t)=P_0(1+\lambda(\varepsilon t)) \\
Q(\varepsilon t)=Q_0(1+\lambda(\varepsilon t))
\end{gathered}
\end{equation}
where $\lambda$ is the loading factor which gradually increases. The maximum value $\lambda P_0$ before the SNB occurs corresponds to the voltage stability margin. 

As a result, the stochastic power system model (\ref{eq:sde_model_u}) can be represented by:
\begin{equation}
\label{eq:sde_model_uee}
\dot{\bm{u}} = \bm{G}(\bm{u},\bm{p}(\varepsilon t))+\sigma B\bm{\xi} \\
\end{equation}

We aim to study the impact of the intensity of the stochastic perturbation $\sigma$ and the load power increasing speed $\varepsilon$ on SNB of the model (\ref{eq:sde_model_uee}) and thus on the voltage stability margin of power systems. 

Detailed simulation study in Section \ref{4} shows that both the strength of load fluctuation and the load changing speed can affect the size of the voltage stability margin. 

\vspace{-5pt}

\section{Numerical Results}\label{4}
\vspace{-2pt}
Numerical study and Monte Carlo simulations 
were carried out
on the IEEE 14-bus system in which an ERL dynamic load was added to Bus 9.
Random fluctuations were included in the ERL model as in (\ref{eq:exp_rec_load_sde}).
PSAT Toolbox \cite{Milano} was used to perform all the simulations. Euler-Maruyama method was used to generate the Ornstein-Uhlenbeck process and the numerical integration step size was $\Delta t=0.05$s.

\vspace{-5pt}
\subsection{The Impacts of the Load Power Changing Speed on the Voltage Stability Margin}

If there is no uncertainty, i.e., $\sigma=0$ in (\ref{eq:sde_model_withouty_2}), the voltage stability margin is 5.4275 per unit (pu). To investigate how the
increasing speed of load power will affect the voltage stability margin if the randomness of load variation is considered, we increased the real and reactive power of the PQ load at Bus 4 at five different speeds, namely 2\% of its nominal values $P_{0}$ and $Q_{0}$, every 0.1s, 0.5s, 1s, 2s and 5s, while keeping the fluctuation intensity of the ERL load as $\sigma=0.10$.

\begin{figure}[!ht]
\vspace{-3pt}
\centering
\includegraphics[width=2.2in ,keepaspectratio=true,angle=0]{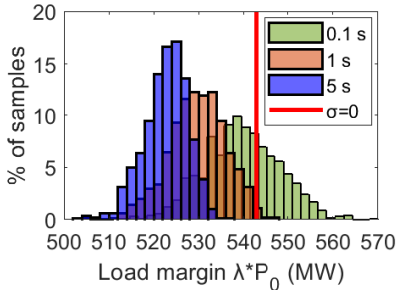}
\vspace{-5pt}
\caption{The distribution of the voltage stability margin for different load increasing speeds in the case where $\sigma=0.1$.}\label{margin_allspeeds}
\vspace{-14pt}
\end{figure}

Table \ref{sigma0.1_all} 
presents the statistics (the mean and the variance in the first two rows) and the 90\% confidence interval (the last two rows)
of the load margin $\lambda P_0$ obtained from 1000 Monte Carlo simulations for each of the five cases.
It can be seen that the size of the stability margin will not be greatly affected if the load changing speed is fast. However, as the load changing speed decreases (i.e., longer time passes before each increase),  $\mathbb{E}(\lambda P_{0})$ starts decreasing. Indeed,  $\mathbb{E}(\lambda P_{0})$  can be decreased up to 3.12\% as the load changing speed decreases.
In addition, Fig. \ref{margin_allspeeds} shows the histograms of the voltage stability margin for the different cases, in which we can see that there is a shift of the curves to the left for slower speeds. %

The above observation makes sense since the system takes longer time to reach the SNB if slower load changing speed is applied, during which the randomness may accumulate and thus it may drive the system to the SNB sooner. These results imply that the time evolution property of the driving parameters needs to be carefully considered to achieve accurate voltage stability assessment due to the randomness brought about by loads. 
It is worth noting that such important results cannot be observed by using deterministic or static approaches.

\vspace{-10pt}
\begin{table}[!ht]
\centering
  \caption{The statistics and the 90\% confidence interval of the voltage stability margin variation for various load increasing speeds}\label{sigma0.1_all}
  \begin{tabular}{|c|c|c|c|c|c|c|}
\hhline{|=|=|=|=|=|=|}
\begin{tabular}{c}Load\\ Increasing \\ Speed (s) \end{tabular}&0.1&0.5&1&2&5\\
  \hline
  $\mathbb{E}(\lambda P_0)$ (pu)&5.40&5.3299&5.2991&5.2678&5.2313 \\
  \hline
  $\mathrm{Var}(\lambda P_0)$&0.0074&0.0052&0.0037&0.0030&0.0023\\
 \hline
 $\mathbb{E}(\lambda P_0)- d$ (pu)&5.3955& 5.3262&5.2959&5.2650&5.2288 \\
   \hline
  $\mathbb{E}(\lambda P_0)+ d$ (pu)&5.4045&5.3337&5.3022&5.2707&5.2337\\
\hhline{|=|=|=|=|=|=|}
  \end{tabular}
\vspace{-10pt}
\end{table}


\vspace{-5pt}
\subsection{The Impacts of the Load Fluctuation Intensity on the Voltage Stability Margin}

In this part, the impact of the intensity of load fluctuations on the voltage stability margin is studied. Intuitively, the larger the fluctuation intensity is, the smaller the voltage stability margin would be. Nevertheless, validation of such conjecture is lacking in the literature. Therefore, we used Monte Carlo simulations to analyze the probabilistic characteristics of the voltage stability margin under the influence of various load variation intensities. We applied three different intensities, i.e., $\sigma_1=0.05, \sigma_2=0.10$ and $\sigma_3=0.15$, respectively, to model the Ornstein-Uhlenbeck fluctuations for the ERL at Bus 9, while keeping the same changing speed for the PQ load at Bus 4. Particularly, the constant PQ load increased its nominal power by 2\% per second.

\begin{figure}[!h]
\vspace{-2pt}
\centering
\includegraphics[width=2.2in ,keepaspectratio=true,angle=0]{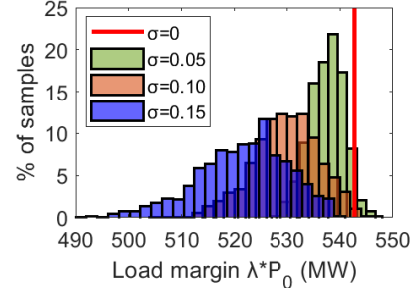}
\vspace{-5pt}
\caption{The distribution of the voltage stability margin for the same load increasing speed yet different noise intensities. 
}\label{largersigma}
 \vspace{-14pt}
\end{figure}

The mean value, 
the variance, and  the 90\% confidence interval 
for the voltage stability margin are reported from 1000 realizations for each case in Table \ref{largersigma_table}. 
From these results, it can be observed that $\mathbb{E}(\lambda P_{0})$  decreases as the load fluctuation strength becomes larger. Particularly, the percentage of the decrease observed is 3.01\% as the fluctuation intensity grows from 0.05 to 0.15.
The distributions of the voltage stability margin for the different load variation intensities are depicted in Fig. \ref{largersigma}, from which we observe a shift of the histograms to the left when it comes to larger $\sigma$. Such results from systematic Monte Carlo simulations are consistent with our previous conjecture.

\begin{figure}[!h]
\centering
  \includegraphics[width=2.7in,keepaspectratio=true,angle=0]{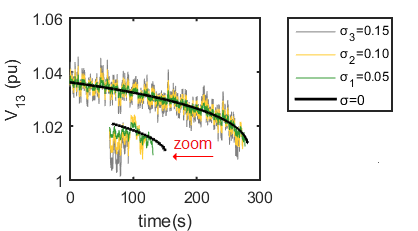}
  \vspace{-5pt}
  \caption{The voltage magnitude at Bus 13 for one realization with different noise intensities.}
  \label{onerealization}
  \vspace{-10pt}
\end{figure}

\begin{figure}[!h]
\vspace{-2pt}
\centering
  \includegraphics[width=2.7in ,keepaspectratio=true,angle=0]{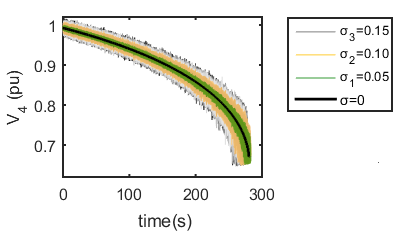}
  \vspace{-5pt}
  \caption{The voltage magnitude at Bus 4 for different noise intensities.}
  \label{allsigmavsdet}
  \vspace{-10pt}
\end{figure}

Additionally,  Fig. \ref{onerealization} presents the voltage magnitude $|V_{13}|$ at Bus 13  for one realization of $\sigma_1, \sigma_2, \sigma_3$. It can be observed that the voltage collapse occurs earlier for larger values of $\sigma$, leading to a smaller voltage stability margin. All the other bus voltage magnitudes exhibit similar behaviors. To have a clearer picture about the statistical characteristics of the trajectories,  Fig. \ref{allsigmavsdet} shows a comparison of the 1000 realizations of $|V_4|$ for different $\sigma$. It corroborates the analytical results shown in Section \ref{sectionconcentrationofpaths} that the depth of the concentration neighborhood $\cal B$$(h)$ depends on the fluctuation intensity $\sigma$. The larger fluctuation intensity is, the wider the trajectory distribution is expected.


\vspace{-5pt}
\begin{table}[!h]
\centering
  \caption{The statistics and the 90\% confidence interval of the voltage stability margin variation for various noise intensities}\label{largersigma_table}
  \begin{tabular}{|c|c|c|c|}
\hhline{|=|=|=|=|}
Fluctuation intensity $\sigma$&0.05&0.10&0.15\\
  \hline
   $\mathbb{E}(\lambda P_0)$ (pu)&5.3768&5.2991&5.2148\\
   \hline
  $\mathrm{Var}(\lambda P_0)$&0.0012&0.0037&0.0070\\
    \hline
   $\mathbb{E}(\lambda P_0)- d$ (pu)&5.3750&5.2959&5.2104\\
   \hline
  $\mathbb{E}(\lambda P_0)+ d$ (pu)&5.3786&5.3023&5.2192\\
\hhline{|=|=|=|=|}
  \end{tabular}
\vspace{-10pt}
\end{table}


\section{Conclusions}\label{5}

This paper investigates the impacts of stochastic load fluctuations on the size of the dynamic voltage stability margin using bifurcation theory. Through systematic numerical study, it has been shown that both the load fluctuation intensity and the load power changing speed may affect the size of the voltage stability margin.
Particularly, it seems to be the first time to reveal the influence of the time evolution property of the driving parameters on the voltage stability margin in the presence of uncertainty. It has been observed that a slower changing speed  of load power or a larger load variation intensity may lead to a smaller voltage stability margin. Therefore, it is crucial to consider 
both factors in order to accurately assess the voltage stability.
It is worth mentioning that such outcomes cannot be observed by using static or deterministic approaches, which in turn reinforces the importance of carrying out dynamic and stochastic approaches in voltage stability analysis, especially considering the increasing degree of uncertainty in modern power systems due to the integration of RES.


\vspace{-6pt}



\bibliography{paper}

\end{document}